\documentclass[11pt]{article}

\usepackage{cite}
\usepackage{amssymb}
\usepackage{amsmath}
\usepackage{bm}

\begin{document}

\title{Aspects of Algebraic Quantum Theory: a Tribute to Hans Primas}
\author{B. J. Hiley\footnote{E-mail address b.hiley@bbk.ac.uk.}.}
\date{Physics Department, University College, London, Gower Street, London WC1E 6BT.\\ TPRU, Birkbeck, University of London, Malet Street, \\London WC1E 7HX. }
\maketitle

\section{Introduction}

\subsection{The Common Ground}

It is a privilege to be invited to contribute to this volume dedicated to Hans Primas whose work on the foundations of quantum theory has had a strong influence on my own thinking on the subject.  I first came across his ideas on algebraic quantum mechanics in a bound manuscript entitled {\em Quantum Mechanical System Theory}\cite{hpum09} in David Bohm's room at Birkbeck College in 1977.  The manuscript, co-authored with Ulrich M\"{u}ller-Herold, was to prove invaluable for my thinking about quantum theory.  

 I had been working with David Bohm trying to develop a new way of thinking about quantum theory based on a process philosophy, in which we were trying to formulate in terms of an algebraic structure along the lines of the original proposals of Born, Heisenberg and Jordan\cite{mbpj25}.
    The idea of using an algebraic structure to describe process has an even longer history going back to Hamilton\cite{wh67}, Grassmann\cite{gg94, gg95} and Clifford\cite{wc82}, 
 but for one reason or another it fell into disrepute, in spite of its use by Eddington\cite{ae36}.  
 
 Fortunately today the 
 notion of process as fundamental is undergoing a revival, particularly with the appearance of category theory especially in the hands of  Abramsky and Coecke\cite{asbc04} and Coecke\cite{bc05} who use the theory in the context of quantum mechanics, explaining in greater detail their motivations for using a process approach.  In this paper I prefer to motivate the algebraic theory along lines that are more closely linked with the approach developed by Primas.
Indeed it was his manuscript that first drew my attention to the advantages of the more general $C^*$-algebraic approach, an algebraic structure that I was completely unaware of at the time.  

My interests in an algebraic approach had already been aroused by Penrose's\cite{rp71} twistor theory, a generalisation of the Dirac Clifford algebra introduced by Dirac to describe the relativistic electron. At the time Penrose was in the mathematics department at Birkbeck and, together with Bohm, we would meet regularly for seminars that were concerned with the possibility of developing quantum space-time structures, a radical idea that we thought necessary in order to unite quantum theory with general relativity.  

Penrose\cite{rp71} was also exploring the possibility of developing a description based on a discrete spin network, thus avoiding the need to assume an {\em a priori} given space-time continuum\cite{rp67}.  
This idea of a network structure fitted in very nicely with the topic of my PhD, although that was in a very different field. 
 
My thesis involved investigating certain aspects of the Ising model used in the study of cooperative phenomena in solid state physics. 
The simple model that I was exploring involved determining the thermodynamics of a many-particle lattice system with nearest neighbour interactions.  It was based on a method of finite clusters, using an idea first proposed by Domb\cite{cdbh62}.
The evaluation of the partition function, and hence the thermodynamical properties, necessitated developing a technique for embedding finite graphs in regular tessellations.  What I noticed was that some of these properties, essentially combinatorial in nature,  depended only on the dimensionality of the embedding space and not on the detailed structure of the tessellation. In other words, simply by {\em counting} embeddings, one could determine the dimensionality  of the embedding space\cite{bhaf77}. 
 It was only later that I became aware of the fact that the partition function could be obtained much more simply using an  algebraic approach used in knot theory.  This approach is described in Kauffman\cite{lk01}  who illustrated  the technique on small clusters.

The phrase `quantum space-time' was a generic term to refer to any structure that did not take a continuum of points as fundamental, but rather the points were assumed to emerge from a deeper structure.  That was, in fact, the idea behind the Penrose twistor which is used to describe a complex of light rays whose  intersections define the points of space-time.  He also found that congruences of light rays twisting around each other could be used to define sets of `extended points' which he hoped would avoid some of the singularities that plague quantum electrodynamics.

 But surely finding partition functions of a spin lattice is a long way from the problems of developing a quantum space-time?  Not so because it turns out that the algebraic techniques lying behind both twistors and the algebraic evaluation of partition functions are closely related to the seminal work of Vaughn Jones\cite{vj86} on von Neumann algebras.  In a remarkable paper, he showed the connection between these 
 algebras and the combinatorial properties of knots which, as we have already remarked, lie at the heart of the techniques involved in evaluating the partition function of finite clusters of spin systems. The connection becomes even more suggestive when it is realised that the Onsager exact solution\cite{lo44}
 for the two-dimensional Ising model involves a Clifford algebra, an algebra that is one example of a von Neumann algebra.  Note also that these algebras are the very algebras that Penrose\cite{rp71} used to construct his twistors.   However all these ideas were then yet to unfold in the future.
 
 \subsection{Structure-Process}

In those early days,  Bohm\cite{db65, db71} was developing his notions of ``structure-process" which emphasised the relationships, order and structure of a network of elementary processes. Not 
relations that could be embedded in the Cartesian order of points, but a new order from which the classical Cartesian order could be abstracted in some suitable limit.  This structure, we believed, would provide a more natural way of accounting for quantum phenomena.

The basic ideas of `structure' had already been introduced by Eddington\cite{ae58} when he raised the question ``What sort of thing is it that I know"?
For him the answer was {\em structure}, structure that could be captured by mathematics.  For example, the concept of space is not an empty `container',  but a relationship of the ensemble of movements that is experienced  as we probe our surroundings, using light signals or other suitable physical processes.
 For Eddington, the structure of these experiences could be captured by a group, which in the relativistic case would be the Lorentz group, giving rise to Minkowski space-time.  Of course in the presence of a gravitational field, this group must be replaced by a larger group, the group of general coordinate transformations  but for Penrose the conformal group was general enough to be explored initially.   When we come to quantum phenomena, Weyl\cite{hw31} pointed out that we must turn our attention, not to the group, but to the group algebra.
It is the group algebras that  form the background to non-commutative geometry.

\subsection{The Role of Clifford Algebras}

 However, again,  I go too fast because initially Bohm and I thought that a natural mathematical expression of this structure would be provided by combinatorial topology alone\cite{dbbh70}.  Although this provided 
 some interesting insights, it misses a vital ingredient, namely, the activity or movement that was necessary to describe process. But then I noticed that Penrose's spin network had Clifford algebras at its heart, the algebra that Onsager used to solve the two-dimensional Ising model.  Could it be that the combinatorial aspects could be captured by an algebra itself, so that we   could use algebras to describe a dynamic structure-process?

 To my surprise I found that Clifford\cite{wc82} was led to his algebra, not  by thinking of a quantum system, but by considering the dynamical activity of classical mechanical systems. He noticed that Hamilton's quaternion algebra, a way of describing rotations in space through action, could be generalised to capture the  Lorentz group and even leads to  the conformal group which is used in twistor theory.  Algebraic elements could be understood in terms of how movements could be combined to form new movements. Clifford
 introduced terms like `versors', `rotators', and `motors' emphasising activity. 
Unfortunately these ideas seemed to add nothing new to physics that was not already described more simply by the vector calculus, so the algebraic approach was ignored. However that changed when Dirac, faced with the negative energies appearing in the relativistic generalisation of the Schr\"{o}dinger equation, rediscovered the Clifford algebra.  It provided a description of spin, relativity and the twistor in one algebraic hierarchy.
 
 Unfortunately the appearance of the Dirac Clifford algebra did not lead to a reconsideration of Clifford's ideas.  Rather the algebra was seen as a generalisation of the quantum operator algebra that was already used in the standard Hilbert space formalism taught to undergraduates.  In that approach the wave function played a key role and gave rise to the so called `wave-particle duality', a notion that I find very unhelpful, being a totally confused idea.  Somehow this wave function is used to describe the so called `state of the system' which was, in turn, assumed to evolve in the Cartesian order of space and time.  While this approach was a predictive success, it has many, as yet, unsolved interpretational problems, such as the measurement problem, schizophrenic cats and the like. All of these could be handled as a set of rules for getting `correct' results, but one is left with the uneasy feeling that something is not quite right because the nature of  the physical processes themselves remains very unclear.
 
 This view was shared by Hans Primas who proposed the question ``Why a Hilbert space model?"  He then explained that this Hilbert space was but a particular representation of a more general quantum mechanics.  The algebra emphasises a non-commutative structure, a structure that has its  origins in the early work of Born, Heisenberg and Jordan\cite{mbpj25}.  
For Primas\cite{hpum09b} 
 \begin{quote}
 Algebraic quantum mechanics starts with an abstract $B^*$-algebra, ${\cal A}$,  of observables. From this algebraic realisation of quantum mechanics, we can get the corresponding Hilbert-space model ${\cal H}$ .... as the universal representation $(\pi, {\cal H})$ of the $B^*$-algebra ${\cal A}$.
 \end{quote}
Thus Hilbert space is a mere representation, but a representation of what?  Could algebraic structure itself provide a description of structure-process and in doing so, clarify the nature of quantum processes?
 
 \section{The Propositional Calculus and  Algebraic Idempotents}
 
\subsection{von Neumann Algebras and a Propositional Calculus}

We now come to the point where algebra meets logic.  Primas highlighted the close relationship between the von Neumann algebras and orthomodular lattices of the type used in the analysis of formal logic.  In fact the set of projections in a von Neumann algebra forms a complete orthomodular lattice so that investigating the properties of this lattice gives a different insight into the algebraic structure.

Projection operators are idempotents, $E^2=E$ and because their eigenvalues are $0$ and $1$,  they can be used to define the truth or falsity of a set of propositions.  We thus have an alternative method of analysing the Schr\"{o}dinger formalism in terms of a non-Boolean logic, a generalisation of the Boolean logic of classical physics. 

 The generalised non-Boolean logic contains a new notion of {\em incompatible propositions}, tied intimately to the appearance of non-commuting operators.  This difference led Finkelstein\cite{df68} to conclude that the appearance of quantum processes causes Òa fracture in physical logicÓ.  Indeed  Finkelstein showed that in this non-Boolean logic, the distributivity law of classical logic was violated.

This raises the important question as to whether this change in logic has to do with the fact that we can only obtain {\em incomplete knowledge} of a quantum system or whether this fact stems from a profound change in the basic {\em reality} underlying quantum phenomena.  Bohr offered an epistemological interpretation in which he proposed that the incompatibility of propositions arises from our inability, {\em  in principle}, to obtain complete knowledge of the system.  For Bohr, quantum phenomena confirmed that there was  a new principle of epistemology, namely the principle of complementarity to which all knowledge must conform.  If this was a fundamental  principle then, no matter what underlies appearance, it would be impossible, even in principle, to construct intuitive pictures of this underlying reality, pictures of the type used in the classical world.

However quantum phenomena occur without the need for anyone to {\em interpret} them or have knowledge of them. There is an actual process unfolding, independent of any observer and this fact demands an underlying ontology.  As Primas\cite{hp77} insists 
\begin{quote}
Accordingly, practically all high-level theories adopt some kind of scientific realism i.e. the view that biological, chemical and physical objects have existence independent of some mind perceiving them.
\end{quote}
The key question is then, ``What form is this ontology going to take"?  Is it going to be `veiled reality' as suggested by d'Espagnat\cite{be03} 
or do we follow Primas\cite{hp77} and insist that ``the most fundamental theory has to be phrased in an {\em individual and ontic interpretation}".
Our hope was that the notion of structure-process would provide the intuitive basis of such a fundamental theory. 

 Any generalised theory must be based on
 non-commutative algebras that lie at the heart of quantum processes.  Since geometry forms the basis of classical physics, its generalisation, non-commutative geometry, must be the way forward to explore the nature of the underlying ontology.

Such a  possibility had already been anticipated by Murray and von Neumann \cite{vnm36}, who presented a very detailed, but intimidating mathematical discussion 
of  what are now called von Neumann algebras, algebras that would play a fundamental role in non-commutative geometry\cite{mk09}.
Fortunately for the purposes of this paper we will not require this detailed knowledge as we can illustrate the essential ideas using the orthogonal Clifford algebra, a specific von Neumann algebra but one with which  physicists and chemists are very familiar through the use of the Pauli $\sigma$-matrices and the Dirac $\gamma$-matrices.

What the physicist or the chemist may not realise, however,  is that a Clifford algebra over a complex field is a particular example of a type $II_1$ von Neumann algebra with a Jones index of $4\cos^2(\pi/4)$\cite{vj03}.  From the comments above, it should be clear that the Clifford algebra will play an important role in our discussion of a non-commutative geometry,
a point of view shared by Finkelstein\cite{df87} 
when he writes, ``I am strongly tempted by the example of Clifford".

\subsection{The Role of the Clifford Algebra in Non-commutative Quantum Geometry}
  
 As we have already remarked, the conventional view among physicists is to regard the Clifford algebra  merely as a formal mathematical device, but our introductory remarks suggest that it is more than that, describing an underlying structure-process.  However to proceed down that route means we must give up, as a fundamental form, the classical notion of a particle evolving along a well defined trajectory in an  {\em a priori} given space-time.  Instead we should adopt a thoroughgoing process philosophy along the lines suggested by Eddington\cite{ae58}, Finkelstein\cite{df96} and Bohm\cite{db80}.  

\subsection{What are Quantum Particles?}

To summarise then, in a process philosophy, we must give up the common sense idea that the world consists of material objects with definite size, shape and properties.  But this notion has already been called into question in special relativity where we are forced to adopt a description based on the notion of a point event.  There is no consistent description of an extended rigid object; a particle must be treated as a complex structure of events that can be regarded as forming a `world tube'. The tube itself cannot have a sharp boundary but must be identified with a pattern of events, distinguishable, but not separate from, a complex of interrelated background events.  In this approach the `particle' is a semi-stable, quasi-local feature that can preserve its form in time.  However under suitable conditions it can undergo, not only quantitative changes, but also qualitative changes, in its basic elements, a phenomenon that is well-known in high energy particle physics.  

In passing, note that Primas\cite{hp77} also has a similar structural notion of a  `particle'.  He stresses  that the so-called `fundamental' entities, such as electrons, protons, or quarks, must not be taken as the building blocks of reality.  They are merely what he calls  {\em patterns of reality}.  For Primas these patterns emerge operationally from the empirical domain, a point to which I will return later.

A limitation of the notion of an elementary `rock-like' particle becomes even more apparent in the quantum domain. To bring the difficulty out clearly, consider the following example inspired by Weyl.  Suppose we retain the classical notion of a particle with specific properties.  To keep things simple, consider a quantum world in which we have a collection of objects with two distinct shapes, either spheres or cubes, and two distinct colours, either red or blue.  Our task is to separate these objects into four distinct groups -- red spheres, blue spheres, red cubes and blue cubes.  In a classical world there is no problem, but in this quantum world, 
 shape and colour are observables, represented by non-commuting operators, their `values' being represented by their corresponding eigenvalues.  This means that to separate colours and shapes, we must have two different types of observing instruments.  In our case we call these instruments `spectacles'.
 
Suppose we require to collect together an ensemble of red spheres. First we put on the `shape-distinguishing' spectacles and collect together spheres, discarding all the cubes.  Then we put on the `colour-distinguishing' spectacles and collect together the red spheres, discarding all the rest.  We are done; we have a collection of red spheres.  So what is the problem?   Just recheck that the objects in the ensemble are still spheres.  We use the first pair of glasses again and find that half the objects are now cubes! No permanent {\em either/or} in this world.  No permanent {\em and/and} either!  

Clearly quantum phenomena do not have their existence defined in terms of classical objects with well defined properties!   Finkelstein has already stressed this feature and argues that  ``to speak about the wave function of the system is a syntactic error"\cite{df87}.
We do not simply `find' the state of a system.  We have to `probe' the system with another physical process, the `observing instrument'.  In other words our instruments are part of the underlying structure-process and therefore change the system itself, or better still, change the process that {\em is} the system. How, then, do we encompass these radically new ideas without losing features of the standard formalism that have been used with outstanding success?

Let us begin by following Eddington \cite{ae58} who suggests that the {\em elements of existence}, the individuals, in a process world, should be described by idempotents, $E^2=E$.   The eigenvalues, $\lambda_{e}$, of an idempotent are 1 or 0, existence or non-existence.  In symbols
\begin{eqnarray*}
E^{2} = E, \hspace{1cm} 	\mbox{with}  \hspace{1cm} 	\lambda_{e} = 1 \; \mbox{or}\; 0.
\end{eqnarray*}
If all idempotents commute, as in classical physics,  existence is always well defined. We have a Boolean logic.  In quantum theory we have a difference, idempotents do not always commute 
\begin{eqnarray*}
[E_{a}, E_{b}] \neq 0.	
\end{eqnarray*}
What then of existence?
\begin{eqnarray*}
\mbox{Either}\; E_{a} \;\mbox{or}\; E_{b},\hspace{0.5cm}\mbox{never}\hspace{0.5cm}		E_{a} \;\mbox{and}\; E_{b}.	
\end{eqnarray*}
Existence, non-existence and in between?  This is the consequence of a non-Boolean logic.

\subsection{Idempotents and Clifford Algebras}

The suggestion is that the idempotent will provide a means of focusing on the {\em sub-process} that {\em is} the individual.  The individual is a process that is continually changing into itself, $E.E=E$. While probing the individual,  the process may change the quality of the idempotent, it nevertheless remains an idempotent, enabling us to track the individual as a  sub-process within the whole structure-process.  In an algebra, an idempotent  can be used to define a set of elements within a minimal left ideal of the total algebra.  These elements carry all the information contained in the `wave' function but now have the advantage of being an integral part of the whole algebra.

  In a semi-simple algebra, we can always form an element of such an ideal by writing $\Psi_L({\cal A})=\psi_L({\cal A})E$.  Mathematically we are constructing a left module or left vector space, but we need not be familiar with this mathematical structure to see how it works.
Consider a spin-half system which requires the observables to be expressed in terms of the Pauli spin matrices.  As is well known the spin `wave' function is a column two-matrix, the spinor, 
\begin{eqnarray*}
\Psi=   \begin{pmatrix} 
      \psi_1 \\
      \psi_2 \\
   \end{pmatrix}.
\end{eqnarray*}
From the algebraic point of view, the Pauli spin matrices define the Clifford algebra $C_{3,0}(\sigma)$ generated by the three Pauli spin matrices $\sigma_i$.  An element of a minimal left ideal can be written in the form $\Psi_L(\sigma)=\psi_L(\sigma)E$ where $E$ is some idempotent. It is conventional to choose $E=(1+\sigma_3)/2$, which breaks the rotational symmetry and defines a preferred $z$-axis while  $\psi_L(\sigma)\in{\cal A}$.  

If we then polar decompose the algebraic spinor, we can write $\Psi_L(\sigma)=RU$ where $U=U^\dag$ and $R$ is a positive definite matrix. 
It is then easy to show that the spinor can be written in the form 
\begin{eqnarray*}
\Psi_L(\sigma)=g_0+g_1\sigma_{23}+g_2\sigma_{13}+g_3\sigma_{12};\quad\quad g_i\in\mathbb R.
\end{eqnarray*}
Here we have written the elements of the algebra in terms of Pauli matrices, $\sigma_{ij}=\sigma_i\sigma_j$, a rotor.
To make contact with the usual spinor, we have the identities 
\begin{eqnarray}
g_{0}= (\psi_{1}^{*}+\psi_{1})/2\hspace{0.5cm}g_{1}=i(\psi_{2}^{*}-\psi_{2})/2 \nonumber\\ g_{2}=(\psi_{2}^{*}+\psi_{2})/2
\hspace{0.5cm}g_{3}=i(\psi_{1}^{*}-\psi_{1})/2.	
				\label{eq:g-psi}			
\end{eqnarray}

Let us emphasise again that we have chosen a specific idempotent, namely, $E=(1+\sigma_3)/2$ which means that we have broken the spherical symmetry by picking a specific direction, conventionally the $z$-axis.  This is usually done by introducing a homogeneous magnetic field, so the choice of idempotent is defined {\em operationally}, just as Primas' patterns are defined operationally.  In other words we are changing the process that {\em is} the system under investigation.  In Wheeler's words\cite{jw91}, we are participating in the process to induce a change in the process that constitutes the system.  

This is exactly what we need to account for our toy model of a quantum world using `shapes' and `colours'.  The change that we find when checking the content of the final ensemble arises from the participatory nature of our `instrument'.  Looking through the `quantum  spectacles' is not a passive process, it is an {\em action}, which must not be thought of as a mere `disturbance'.  It is an inescapable change in the structure-process that {\em is} the system.
More details of this idea will be found in Hiley and Frescura\cite{bhff80} and in Hiley and Callaghan\cite{bhbc10}.

This example explains very succinctly how the Pauli algebraic spinor appears and is used in the  description of the algebra.  It is easily generalised to the Dirac spinor and indeed the twistor, which is a semi-spinor of the conformal Clifford.  These Clifford algebras form a hierarchy or tower of algebras,  $C_{3,0}\rightarrow C_{1,3}\rightarrow C_{4,1}\rightarrow C_{2,4}$ of the type considered by Jones\cite{vj86}.  It is interesting to note that the Schr\"{o}dinger `wave' function can also be considered as an element of a minimal left ideal in the Clifford algebra $C_{0,1}$, with the quaternions appearing in $C_{0,2}$.

In addition to elements of the left ideal, we also have dual elements,  $\Psi_R({\cal A})=E\psi_R({\cal A})$, chosen from an appropriate minimal  right ideal.  This enables us to give a complete specification of the structure-process of an individual system by writing
\begin{eqnarray*}
\rho_c({\cal A})=\Psi_L({\cal A})\widetilde\Psi_L({\cal A})
\end{eqnarray*}
where $\rho_c({\cal A})$ is an element that characterises the system.  It is the algebraic analogue of the density matrix.

If we define $\widetilde\Psi_L({\cal A})=\Psi_R({\cal A})=E\widetilde\psi_L({\cal A})$ then, by suitable choice of the tilde operation, we find $\rho_c^2=\rho_c$, a signature of what is known in the standard approach as a {\em pure state}.  It should be noted that the corresponding dual element  introduced by Primas and M\"{u}ller-Herold\cite{hpum09} was called a {\em normalised positive linear functional}.  Using this additional mathematical structure, we have the  possibility of a generalisation to mixed states, but in this paper,  we confine our attention to pure states for simplicity.

As well as rotational symmetries, we must also consider translation symmetries, which implies turning our attention to the Heisenberg algebra.  Here there is a technical problem because this algebra is nilpotent and therefore does not contain any idempotents.  However Sch\"{o}nberg\cite{ms57}, and later Hiley\cite{bh01}, showed that it was possible to 
extend this algebra by adding sets of idempotents to form a symplectic Clifford algebra\cite{ac90}. This then enables us to employ 
similar techniques to those used in the orthogonal Clifford algebra.  One is then able to find time development equations that correspond to the  Heisenberg equations of motion.

The characteristic element $\rho_c({\cal A})$ can now be subjected to both left and right translations to determine two fundamental time development equations,
\begin{eqnarray}
i[(\partial_{t}\Phi_{L}){\widetilde\Phi_{L}}+\Phi_{L}(\partial_{t}{\widetilde\Phi_{L}})]=i\partial_{t}\rho_{c}=(\overrightarrow{H}\Phi_{L}){\widetilde\Phi_{L}}-\Phi_{L}({\widetilde\Phi_{L}}\overleftarrow{H})
				\label{eq:conprob}
\end{eqnarray}							
and
\begin{eqnarray}
i[(\partial_{t}\Phi_{L}){\widetilde\Phi_{L}}-\Phi_{L}(\partial_{t}{\widetilde\Phi_{L}})]=(\overrightarrow{H}\Phi_{L}){\widetilde\Phi_{L}}+\Phi_{L}({\widetilde\Phi_{L}}\overleftarrow{H}).
					\label{eq:anticom}
\end{eqnarray}			
We now have the possibility of two forms of Hamiltonian  $\overrightarrow{H}=\overrightarrow{H}(\overrightarrow{D}, V, m)$ and $\overleftarrow{H}=\overleftarrow{H}(\overleftarrow{D}, V, m)$ emphasising the distinction between left and right translations.
We will not derive these equations here as they have been derived in Hiley and Callaghan\cite{bhbc10}; nevertheless we will use them in the next section.
We merely note that equation (\ref{eq:conprob}) is the quantum Liouville equation expressing the conservation of probability, while equation (\ref{eq:anticom}) is the quantum Hamilton-Jacobi [QHJ] equation expressing the conservation of energy.  A detailed discussion of these equations will be found in Hiley\cite{bh15}.

\section{The Implicate and Explicate Order}

We must now return to discuss the relation between the non-Boolean structure and its Boolean substructures.  Primas\cite{hp77} offers a formal way to understand the relationship between these two logics in terms of a specific physical process.  We will explain his position in the following way.

We have argued that there is no such thing as a direct, faithful observation in a quantum process.  However as Bohr has pointed out,  the results of any observation must be unambiguously described in terms of a Boolean structure.  This is the only way we can unambiguously communicate  the results of an experiment.  How then do we understand the Boolean aspects of a fundamentally non-Boolean process?   

Primas suggests that the results of an experiment can be understood as a {\em pattern} that is formed by detaching ourselves, and our instruments, from properties that we consider to be non-essential. He calls the total process, the {\em factual} domain ${\cal F}_\alpha$, which he distinguishes from the empirical domain ${\cal E}_\alpha$ defined {\em operationally} as the result of the $\alpha$-th pattern recognition technique.  The factual domain is non-Boolean and a-local, while the empirical domain is a Boolean and local structure. 
The link between theory and experiment is then regarded as a mapping ${\cal F}_\alpha\rightarrow{\cal E}_\alpha $ which is not required to be one-to-one.  

Bohm\cite{db80} has made, in essence, a similar proposal to understand the relation between Boolean and non-Boolean aspects of physical processes, but in terms of a more general language.  Structure-process is defined in terms of an algebra in which the individual elements of the algebra, like words, take their implicit meaning from the way in which the algebra as a whole is used.  For example the symbols in the Pauli Clifford algebra take their meaning from the rotational symmetries we experience as we rotate in space.  

In such a structure, all the spin components cannot be made explicit by the same action.  The spin in the $z$-direction can be made explicit, while the other components remain implicit. More generally, as is well known,  an ensemble of properties corresponding to mutually commuting observables can be made explicit together.  This subset of elements  forms  a Boolean substructure within the more general non-Boolean structure.
Bohm called these substructures explicate orders, while the total non-Boolean structure was called the implicate order.

I have used examples from gestalt psychology as a metaphor to illustrate the notions of  the  implicate and explicate order. The young lady/old lady gestalt illustrates succinctly what is involved. Our perception constructs or `explicates' a Boolean pattern, say the young lady, by ignoring some of the details in the drawing.  When none of the details are ignored, we have a non-Boolean structure.  However metaphors are limited and a deeper analysis based on equation (\ref{eq:anticom}) shows that a projection actually creates the explicate order.  It creates a Boolean substructure within the non-Boolean totality.

To see how the projection comes in, let us write the equations (\ref{eq:conprob}) and (\ref{eq:anticom}) in a more familiar notation
\begin{eqnarray}
i\partial\rho=(H|\phi\rangle)\langle\phi|-|\phi\rangle(\phi|H) \label{eq:conprobket}
\end{eqnarray}
and
\begin{eqnarray}
i[(\partial_t|\phi\rangle)\langle\phi|-|\phi\rangle(\partial_t\langle\phi|)]=(H|\phi\rangle)\langle\phi|+|\phi\rangle(\phi|H).    \label{eq:anticomket}
\end{eqnarray}
Now introduce the projection operator $P_a=|a\rangle\langle a|$ and take the trace so that  equation (\ref{eq:conprobket}) becomes
\begin{eqnarray}
\frac{\partial P(a)}{\partial t}+\langle[\rho_c,H]_-\rangle_a=0	\label{eq:aconprob}
\end{eqnarray}
while equation (\ref{eq:anticomket}) becomes
\begin{eqnarray}
2P(a)\frac{\partial S_a}{\partial t}+\langle[\rho_c,H]_+\rangle_a=0.		\label{eq:aanticom}
\end{eqnarray}
To bring out what this means, let us choose an harmonic oscillator Hamiltonian $\hat H=\hat p^2/2m+K\hat x^2/2$ and choose the projection operator $P_x=|x\rangle\langle x|$ so that equation (\ref{eq:aconprob}) becomes
\begin{eqnarray*}
\frac{\partial P_x}{\partial t}+\nabla_x.\left(P_x\frac{\nabla_x S_x}{m}\right)=0.
\end{eqnarray*}
This is just the equation for the conservation of probability in position space.
Using the same procedure on equation (\ref{eq:aanticom}) finally gives us\begin{eqnarray*}
\frac{\partial S_x}{\partial t}+\frac{1}{2m}\left(\frac{\partial S_x}{\partial x}\right)^2-\frac{1}{2mR_x}\left(\frac{\partial^2R_x}{\partial x^2}\right )+\frac{Kx^2}{2}=0
\end{eqnarray*}
which is just the quantum Hamilton-Jacobi equation for the harmonic oscillator.  The QHJ equation is simply the equation Bohm obtains by taking the real part of the Schr\"{o}dinger equation under polar decomposition of the wave function.  This equation contains the quantum potential
\begin{eqnarray}
Q=-\frac{1}{2mR_x}\left(\frac{\partial^2R_x}{\partial x^2}\right ).	\label{eq:Q}
\end{eqnarray}
Notice that this potential does not appear in the algebraic equation (\ref{eq:anticom}) which we are regarding as a description of the implicate order.  It only appears in the projected space.  This space is a Boolean phase space constructed with $(x,p_B(x))$ where $p_B(x)$ is the Bohm or local momentum.  It is in this phase space that trajectories have been constructed by Philippidis, Dewdney and Hiley\cite{cpcd79}.  Thus we 
have constructed a Boolean explicate order.

We could choose another projection operator $P_p=|p\rangle\langle p|$ so that the two equations (\ref{eq:conprob}) and (\ref{eq:anticom}) now become 
\begin{eqnarray*}
\frac{\partial P_p}{\partial t}+\nabla_p.\left(P_p\frac{\nabla_p S_p}{m}\right)=0
\end{eqnarray*}
and
\begin{eqnarray*}
\frac{\partial S_{p}}{\partial t} + \frac {p^{2}}{2m} +
\frac{K}{2}\left(\frac{\partial S_p}{\partial p}\right)^2 - \frac{K}{2R_{p}}\left(\frac{\partial ^{2}R_{p}}{\partial
p^{2}}\right) = 0.
\end{eqnarray*}
This enables us to project out another Boolean phase space based, this time, on $(x_B(p), p)$ where $x_B(p) =-\left(\frac{\partial S_p}{\partial p}\right)$.  Thus using the momentum representation we have constructed another explicate order and thereby revealed  $x,p$ symmetry -- a symmetry that Heisenberg\cite{wh58} claimed was {\em not} present in the Bohm approach.

Bohm chose the $x$-representation as a preferred representation simply because he saw a problem in representing the Coulomb potential in the $p$-representation.  However for other potentials there is no difficulty.  Indeed Brown and Hiley\cite{mbbh00}
showed how the approach worked in the particular case of a cubic potential.

Another criticism that is often made of the Bohm approach is that it does not work for the relativistic Dirac particle.  However Hiley and Callaghan\cite{bhbc12} have shown that we can obtain Lorentz invariant analogues of equations (\ref{eq:conprob}) and (\ref{eq:anticom}) which can then be put into the form of a relativistic QHJ equation.
To do this we need to use the orthogonal Clifford algebra $C_{1,3}$.  The expression for the quantum potential is more complicated but can be shown to reduce to the expression (\ref{eq:Q}) in the non-relativistic limit\cite{bhbc12}.

These examples show what is involved in what Primas calls {\em pattern recognition}.  It is not a `passive' recognition, it actually involves an {\em active construction} of the Boolean pattern.  
 But in doing so new features can be introduced as Primas points out.  In the case of the Boolean phase space considered above, it is the appearance of the quantum potential which can be considered as the appearance of a force.  
 
 This is not unlike the nature of the gravitational force which only appears when we project the curved space-time geodesic to a flat Minkowski space-time.  However there is a significant difference in that the curvature of space-time is universal, whereas the quantum potential is, in a sense, `private', 
 being shared by a group of  {\em entangled} particles.  We could have a situation arising where the quantum potential of one group of entangled particles can be very different from the quantum potential  of another entangled group if the groups are non-interacting but nevertheless share the same region of space-time.  The groups do not experience a common quantum potential, it is not universal since they only experience the quantum potential of their own group.

\section{Conclusion}

In this paper I have given a limited view of a new way of looking at quantum phenomena that Hans Primas was one of the first to draw to our attention.  The disadvantage for pioneers of a new vision is that they do not have access to the later developments, particularly the technical advances, in this case the progress in non-commutative mathematics that has been slowly gathering pace since 1977. However without the initial `struggle' to clear the way, others would not have followed.  I will always be grateful to Hans for his early work and our subsequent discussions which, although at times heated, always provided new insights.

\section*{Acknowledgment}
I should like to thank Glen Dennis  for his suggestions and helpful comments.


\begin{thebibliography}{99}


\bibitem{hpum09} Primas, H. and M\"{u}ller-Herold, U., Quantum Mechanical System Theory: A Unifying Framework for Observations and Stochastic Processes in Quantum Mechanics, {\em Advances in Chemical Physics} {\bf 38}, (1978) 1-107. 

\bibitem{mbpj25}  Born M., Jordan P.,  Zur Quantenmechanik, {\em Z. Phys}. {\bf 34}, (1925)  858-888.

\bibitem{wh67} Hamilton, W. R.,  {\em Mathematical Papers, Vol. 3, Algebra}, Cambridge University Press, Cambridge, 1967.

\bibitem{gg94}  Grassmann, H. G., {\em Gesammeth Math. und Phyk. Werke}, Leipzig,  1894.

\bibitem {gg95}Grassmann, H. G., {\em A New Branch of Mathematics: the Ausdehnungslehre of 1844 and Other Works}, translated by Kannenberg, L. C., Open Court, Chicago, 1995.
 
 \bibitem{wc82} Clifford W.K., {\em Mathematical Papers}, XLII, Further note on biquaternions, 385-94, Macmillan, London, 1882. 

 \bibitem{ae36} Eddington, A. S., {\em Relativity Theory of Protons and Electrons}, Cambridge University Press, Cambridge, 1936.
 
\bibitem{asbc04} Abramsky, S. and Coecke, B., A categorical semantics of quantum protocols.  {\em Logic in Computer Science, 2004}. Proceedings of the 19th Annual IEEE Symposium, 2004.

\bibitem{bc05} Coecke, B., Kindergarten quantum mechanics in {\em Quantum Theory: Reconsiderations of the Foundations III},  pp. 81-98, AIP Press, 2005.

\bibitem{rp71} Penrose, R., Twistor Algebra, {\em J. Maths Phys}., {\bf 8}, (1967), 345-366.

\bibitem{rp67} Penrose, R., Angular Momentum: a Combinatorial approach to Space-time, in {\em Quantum Theory and Beyond}, ed. Bastin, T., Cambridge Uni Press, Cambridge, 151-180, 1971.

\bibitem{cdbh62} Domb, C. and Hiley, B. J.,  On the Method of Yvon in Crystal Statistics,    {\em Proc. Roy. Soc}. {\bf A268} (1962) 506-526.

\bibitem{bhaf77}Hiley, B. J., Burke, T. and Finney, J.,  Self-avoiding Walks on Irregular Structures,   {\em J. Phys}. {\bf A10}, 197-204 (1977).

\bibitem{lk01} Kauffman, L. H., {\em Knots and Physics}, p. 373, World Scientific, Singapore, 2001.

\bibitem{vj86} Jones, V. F. R., A new knot polynomial and von Neumann algebras, {\em Notices of AMS}  {\bf 33} (1986) 219-225.  

\bibitem{lo44} Onsager, L., Crystal statistics. I. A two-dimensional model with an order-disorder transition.  {\em Phys. Rev.} {\bf 65} (1944): 117.

\bibitem{db65} Bohm, D., Space, Time, and the Quantum Theory Understood in Terms of Discrete Structural Process, {\em Proc. Int. Conf. on Elementary Particles}, Kyoto,  (1965) 252-287.

\bibitem{db71} Bohm, D., Space-Time Geometry as an abstraction from Spinor Ordering, in {\em Perspectives in Quantum Theory: Essays in Honour of Alfred Lande}, 78-90, MIT Press, Cambridge, 1971.

\bibitem{ae58} Eddington, A. S.,  {\em The Philosophy of Physical Science}, Ann Arbor Paperback, University of Michigan Press, Michigan 1958.

\bibitem{hw31} Weyl, H., {\em The Theory of Groups and Quantum Mechanics}, Dover, London, 1931.

\bibitem {dbbh70} Bohm, D. J., Hiley, B. J. and  Stuart, A.E.G., On a New Mode of Description in Physics,  {\em  Int. J. Theor. Phys}. {\bf 3}, (1970) 171-183.

\bibitem{hpum09b} Primas, H. and M\"{u}ller-Herold, U., footnote on p. 21, Quantum Mechanical System Theory: A Unifying Framework for Observations and Stochastic Processes in Quantum Mechanics, {\em Advances in Chemical Physics} {\bf 38}, (1978) 1-107. 


\bibitem {df68} Finkelstein, D., Matter, Space and Logic, {\em Boston Studies in the Philosophy of Science}, {\bf V}, (1968) 199-215.



\bibitem{be03} d'Espagnat, B., {\em Veiled Reality: An analysis of present-Day Quantum Mechanical Concepts.} Westview Press, Boulder, Colorado,  2003.

\bibitem{hp77} Primas, H., Theory Reduction and Non-Boolean Theories, {\em J. Math. Biology} {\bf 4} (1977) 281-301.

\bibitem{vnm36}  Murray, F. J., and von Neumann, J., On rings of operators. {\em Annals of Math.} (1936): 116-229.

\bibitem{mk09} Khalkhali, M., {\em Basic Non-commutative Geometry}, EMS Publishing, Zurich, 2009.

\bibitem{vj03} Jones, V. F. (2003).  von Neumann algebras. Lecture Notes from
http://www.math.berkeley.edu/vfr/MATH20909/VonNeumann2009.pdf. 
 

\bibitem{df87} Finkelstein, D.,  All is Flux, in {\em Quantum Implications: Essays in Honour of David Bohm}, Ed by Hiley, B. J. and Peat, D., p. 289-294, Routledge and Kegan Paul, London 1987.

\bibitem {df96}  Finkelstein, D. R., {\em Quantum Relativity: a Synthesis of the ideas of Einstein and Heisenberg}, Springer, Berlin, 1996.

\bibitem{db80} Bohm, D., {\em Wholeness and the Implicate Order}, Routledge, London, 1980.

\bibitem{jw91} Wheeler J. A., {\em At Home in the Universe}, p.286, AIP Press, New York, 1991.


\bibitem{bhff80}  Hiley, B. J. and Frescura, F.A.M.,  The Implicate Order, Algebras and the Spinor,   {\em Found. Phys.} {\bf10}, (1980) 7-31.

\bibitem{bhbc10}  Hiley, B. J. and Callaghan, R. E., The Clifford Algebra approach to Quantum Mechanics A: The Schr\"{o}dinger and Pauli Particles, (2010)  arXiv: 1011.4031.

\bibitem{ms57} Sch\"{o}nberg, M., Quantum Mechanics and Geometry, {\em An. Acad. Brasil. Cien} {\bf 29}, (1957), 473-485.

\bibitem{bh01} Hiley, B. J., in {\em Implications, Scientific aspects of ANPA 22}, 107-121, Cambridge, 2001.

\bibitem{ac90} Crumeyrolle A.,  {\em Orthogonal and Symplectic Clifford Algebras : Spinor Structures}, Kluwer, Dordrecht 1990.

\bibitem{bh15} Hiley, B. J., On the Relationship between the Moyal Algebra and the Quantum Operator Algebra of von Neumann, {\em Journal of Computational Electronics}, {\bf 14} (2015)  869-878.

\bibitem{cpcd79} Philippidis, C., Dewdney, C. and Hiley, B. J.,  Quantum Interference and the Quantum Potential,   {\em Nuovo Cimento} {\bf 52B}, (1979) 15-28.

\bibitem{wh58} Heisenberg, W., {\em Physics and Philosophy: the revolution in modern science}, p. 118, George Allen and Unwin, London, 1958.

\bibitem{mbbh00} Brown, M. R. and Hiley, B. J., Schr\"{o}dinger revisited: the role of Dirac's `standard' ket in the algebraic approach.  Quant-ph/0005026.

\bibitem{bhbc12} Hiley, B. J. and Callaghan, R. E.,  Clifford Algebras and the Dirac-Bohm Quantum Hamilton-Jacobi Equation. {\em Foundations of Physics}, {\bf 42} (2012) 192-208. 

\end{thebibliography}
\end{document}